\documentclass[12pt]{article}

\usepackage{latexsym,cmmib57}
\usepackage{amsmath,amsthm}
\usepackage[psamsfonts]{amsfonts,eucal}
\usepackage{graphicx}
\usepackage{url}
\usepackage{chicago}

\newcommand{\condgen}[6]{{#1}#2 #5 #3 #6 #4}
\newcommand{\bbrd}[1]{\mbox{\rm{I}\kern-.1667em{#1}}}
\newcommand{\EXP}{\mathbb{E}}
\newcommand{\PROB}{\mathbb{P}}

\newcommand{\Probcmd}[2]{\condgen{\PROB}{\Bigl\{}{\Bigm|}{\Bigr\}}{#1}{#2}}

\newsavebox{\fmbox}
\newenvironment{fmpage}[1]
        {\begin{lrbox}{\fmbox}\begin{minipage}{#1}}
        {\end{minipage}\end{lrbox}\fbox{\usebox{\fmbox}}}

\newcounter{algocnt}
\newenvironment{algolist}[1]{%
    \begin{list}{\thealgocnt}
    {\parsep 0in\usecounter{algocnt}\setcounter{algocnt}{0}\renewcommand{\thealgocnt}{{#1}\arabic{algocnt}}
    \setlength{\rightmargin}{0in}
    \settowidth{\leftmargin}{{#1}999}\addtolength{\leftmargin}{\labelsep}}}{\end{list}}
\newcommand{\algotab}[0]{\hspace*{0.5em}}
\newcommand{\mt}[0]{\algotab\ }
\newcommand{\mtt}[0]{\algotab\ \algotab\ }
\newcommand{\mttt}[0]{\algotab\ \algotab\ \algotab\ }
\newcommand{\mtttt}[0]{\algotab\ \algotab\ \algotab\ \algotab\ }

\newcommand{\Algo}[1]{\textsc{#1}}
\newcommand{\setto}{\leftarrow}

\newtheorem{theorem}{Theorem}
\newtheorem{definition}{Definition}
\newtheorem{lemma}[theorem]{Lemma}

\newcommand{\QED}{\hfill$\Box$}

\newcommand{\tsize}{n}
\newcommand{\asize}{r}
\newcommand{\dsize}{\ell}
\newcommand{\sizecount}[1]{C_{#1}}
\newcommand{\multiplicity}{\nu}
\newcommand{\oszlopok}{\mathcal{S}}
\newcommand{\alphabet}{\mathcal{A}}
\newcommand{\nlabelings}{s}

\newcommand{\chone}{\texttt{1}}
\newcommand{\chzero}{\texttt{0}}

\newcommand{\tree}{T}
\newcommand{\state}{\xi}

\newcommand{\ali}[2]{\begin{smallmatrix}{#1}\\{#2}\end{smallmatrix}}
\newcommand{\ascoresym}{M}
\newcommand{\ascore}[2]{\ascoresym\Big[\ali{#1}{#2}\Bigr]}
\newcommand{\iscoresym}{\Lambda}
\newcommand{\iscore}[2]{\iscoresym\Bigl[\ali{#1}{#2}\Bigr]}
\newcommand{\gapo}[1]{\gamma^{(<)}_{#1}}
\newcommand{\gapc}[1]{\gamma^{(>)}_{#1}}
\newcommand{\gape}[1]{\gamma^{(-)}_{#1}}
\newcommand{\charat}[2]{{#1}[#2]}
\newcommand{\varA}[2]{\mathsf{A}[#1,#2]}
\newcommand{\varGS}[2]{\mathsf{gS}[#1,#2]}
\newcommand{\varGP}[2]{\mathsf{gP}[#1,#2]}
\newcommand{\Sch}[1]{\charat{S}{#1}}
\newcommand{\Pch}[1]{\charat{P}{#1}}
\newcommand{\intronat}[3]{{#1}[#2\colon#3]}

\newcommand{\aaX}{\texttt{x}}
\newcommand{\aagap}{\texttt{-}}
\newcommand{\aaq}{\texttt{*}}

\begin{document}
\title{In search of lost introns}
\author{%
	Mikl\'os Cs\H{ur}\"os\thanks{Department of Computer Science and Operations Research, Universit\'e de Montr\'eal, Qu\'ebec, Canada}, 
	J. Andrew Holey\thanks{Department of Computer Science, Saint John's University and the College of St.~Benedict, Collegeville, Minn., USA},
	Igor B. Rogozin\thanks{National Center for Biotechnology Information, National Library of Medicine, National Institutes of Health, 
		Bethesda, Md., USA}
}

\maketitle

\begin{abstract}
Many fundamental questions 
concerning 
the emergence and subsequent evolution of 
eukaryotic exon-intron organization 
are still unsettled. 
Genome-scale comparative studies,
which can shed light
on crucial aspects of 
eukaryotic evolution,
require adequate computational 
tools.

We describe novel computational methods 
for studying spliceosomal intron evolution.
Our goal is to give a reliable characterization of the dynamics of intron evolution. 
Our algorithmic innovations 
address the identification of
orthologous introns, and 
the likelihood-based analysis of 
intron data.
We discuss a compression method for the evaluation of the likelihood 
function, which is noteworthy for 
phylogenetic likelihood problems in general. 
We prove that after $O(n\ell)$ preprocessing time, subsequent evaluations 
take $O(n\ell/\log\ell)$ time almost surely in the Yule-Harding random model of 
$n$-taxon phylogenies,
where $\ell$ is the input sequence length.

We illustrate the practicality of our methods 
by compiling and analyzing a data set 
involving 18 eukaryotes,  
more than in any other study to date.
The study yields the surprising result 
that ancestral eukaryotes were 
fairly intron-rich. 
For example, the bilaterian ancestor is estimated to have had more than 90\% 
as many introns as vertebrates do now. 

\paragraph{Contact:} csuros AT iro.umontreal.ca
\end{abstract}

\section{Introduction}
Typical eukaryotic protein-coding genes 
contain introns, which are
removed prior to translation. 
Key constituents of the spliceosome, which is the 
RNA-protein complex that performs the intron excision, 
can be traced back~\shortcite{CollinsPenny} to the last common ancestor of 
extant eukaryotes (LECA). 
Even deep-branching lineages~\shortcite{introns.trichomonas,introns.giardia}
have introns.
It is thus almost certain that spliceosomal introns were present in 
LECA. 
Moreover, when comparing distant eukaryotes,
intron positions often agree~\shortcite{Rogozin.data}.
The similarity is likely due more to conservation 
of early introns than to parallel 
gains~\shortcite{RoyGilbert.earlygenes,Sverdlov.parallelgain}.
It is thus compelling to 
use genome-scale comparisons to
study intron evolution in different lineages, and even
to estimate the exon-intron organization in 
extinct ancestors.
One of the first such studies, by
\shortciteN{Rogozin.data}, involved orthologous gene sets in eight eukaryotes.
The same data set was reanalyzed by different 
authors~\shortcite{RoyGilbert.earlygenes,csuros.scenarios,Carmel.EM,Nguyen.introns},
using novel methods developed for intron data. 
Subsequent inquiries \shortcite{introns.fungi,Roy.apicomplexa,Roy.plants,Coulombe.mammals}
attest to a renewed interest in understanding 
the specifics of intron evolution within different eukaryotic lineages.
This paper introduces novel computational techniques 
for the analysis of spliceosomal intron evolution,
anticipating more large-scale studies to come.

Section~\ref{sec:identification} describes an alignment 
method for intron-annotated protein sequences, 
as well as a 
segmentation method for identifying 
conserved portions of a multiple alignment. 
Section~\ref{sec:likelihood}
describes a likelihood framework in which 
intron evolution can be analyzed in a theoretically sound manner.
Section~\ref{sec:compress} scrutinizes a compression technique that 
accelerates the evaluation of the likelihood function. 
The compression involves an $O(\tsize\dsize)$-time 
preprocessing step for~$\dsize$ sites and a phylogeny 
with~$\tsize$ species. 
We show that the subsequent evaluation takes 
sublinear, $O(\tsize\dsize/\log\dsize)$ time 
almost surely in the Yule-Harding model of random phylogenies,
even in the case of arbitrary, constant-size alphabets. 
Fast evaluation is 
particularly important when the likelihood is maximized 
in a numerical procedure that computes the likelihood function 
with many different parameter settings. 
Section~\ref{sec:applications}
describes two applications 
of our methods. 
In one application, intron-aware 
alignment was used to validate 
some unexpected features of
intron evolution before LECA. In a second application, 
we analyzed intron evolution in 18 eukaryotic species.
We found
evidence of intron-rich early eukaryotes and a prevalence 
of intron loss over intron gain in 
recent times. 

\section{Identification of orthologous introns}\label{sec:identification}
\subsection{Intron-aware alignment}
Orthologous introns can be identified 
by using whole genome alignments in the case of closely related
genomes~\shortcite{Coulombe.mammals}. 
For distant eukaryotes, however, 
intron orthology can only be established 
through protein alignments~\shortcite{Rogozin.methods}. 
The usual procedure is to 
project intron positions onto 
an alignment of multiple 
orthologous proteins. 
If introns in different species are projected to 
the same alignment position, then the introns
are assumed to be orthologous. 

\begin{figure}
\centerline{\includegraphics[width=\columnwidth]{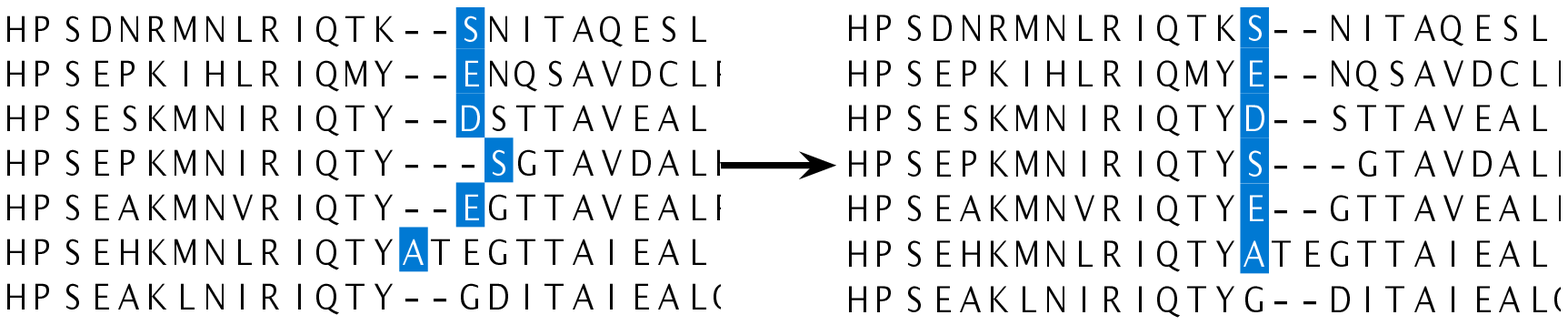}}
\caption{Fragment of a multiple alignment before and after realignment using 
	intron annotation. Shaded rectangles show the intron positions
	projected to the protein sequences.}\label{fig:iali}
\end{figure}

Intron annotation can be included 
in protein alignments
by defining intron match and mismatch scores. 
The alignment score is then computed as the sum 
of scores for amino acid matches, gaps, and intron matches.
Incorporating intron annotation
should lead to better alignments at the amino acid level, and to 
a more reliable identification of 
orthologous introns. 
Figure~\ref{fig:iali} shows 
an example of such improvement.

Intron scoring can be 
based on log-likelihood ratios
in a probabilistic model~\shortcite{Durbin}.
The model is defined by the joint distribution~$p$ 
for the intron state in two sequences~$S$ and~$T$, and 
the prior distributions~$\pi_S$ and~$\pi_T$.
Aligned sites have states~$(s,t)$ with probability $\pi_S(s)\cdot \pi_T(t)$ if 
the two sites are unrelated, 
or with probability~$p(s,t)$ in case of homology.
An $(s,t)$ alignment is scored with 
a value that is proportional to $-\log\frac{p(s,t)}{\pi_S(s)\pi_T(t)}$.

We used the data set of~\shortciteN{Rogozin.data} to 
assess the strength of intron-match signals. Since the data 
include no sites in which all species lack introns,
but the model does allow for that, we 
added extra sites with no introns. 
The original data comprise 7236 intron sites in 684 genes, across 8 species. 
Using a method described earlier \shortcite{csuros.scenarios}, 
we added 35000 unobserved intron sites. 
Using estimates for~$p$ and~$\pi$, we computed the appropriate scores.  
The intron score is asymmetric and varies with 
evolutionary distance and intron conservation. 
Matches for absent introns have an insignificant score, 
but shared introns have a high score, such 
as 93 (human-Plasmodium), 106 (human-Arabidopsis),  
152 (human-{\em S.~pombe}) or 303 (Drosophila-Anopheles)
on a 1/60-bit scale. 
Shared introns thus give 
a signal comparable to amino acid matches: 
in the 1/60-bit scaled version of the VTML240 
matrix~\shortcite{VTML}, 
a tryptophan match scores~289, and an arginine identity scores~113.

Consider the case of aligning two protein sequences, $S$ and~$T$, which
are annotated with the intron positions. 
Every residue has 
two associated intron sites (after the first and second nucleotides of their codons), 
and there is an intron site between consecutive amino acid positions (phase-0 introns). 
Intron sites may or may not be filled in by introns in either sequence. 
We use the notation for $\intronat{S}{i}{0}$ for the phase-0 site
preceding the codon for amino acid~$i$, and $\intronat{S}{i}{1}$, $\intronat{S}{i}{2}$
for phase-1 and -2 sites within the codon.  
Intron presence is encoded by~$\chone$, 
and intron absence is encoded by~$\chzero$ throughout the paper.
The intron annotation is specified by the variables
$\intronat{S}{i}{j}\in\{\chone,\chzero\}$. 
(There can be no introns after the last amino acid.)
Scores for aligned introns
are specified by a~$2\times2$ scoring matrix~$\iscoresym$. 

Phase-1 and phase-2 intron sites are automatically 
placed by their associated amino acids. 
If~$\ascoresym$ is the amino acid scoring matrix,
then the alignment of $S[i]$ and~$T[j]$ entails a score 
of
$\ascore{\Sch{i}}{\charat{T}{j}}+\iscore{\Sch{i}}{\charat{T}{j}}$ 
with
$\iscore{\Sch{i}}{\charat{T}{j}} = \iscore{\intronat{S}{i}{1}}{\intronat{T}{j}{1}}+\iscore{\intronat{S}{i}{2}}{\intronat{T}{j}{2}}$.
Similarly, aligning $S[i]$ with an indel implies a score of
$\ascore{\Sch{i}}{\aagap}+\iscore{\Sch{i}}{\chzero}=\ascore{\Sch{i}}{\aagap}+\iscore{\intronat{S}{i}{1}}{\chzero}+\iscore{\intronat{S}{i}{2}}{\chzero}$,
in addition to possible gap opening and closing penalties. 
Standard alignment procedures need to be modified to deal with phase-0 introns, 
since the placement of phase-0 introns is not fixed with respect to gaps.
It is not possible to simply add a new character to the alphabet to 
represent phase-0 introns because they affect 
gaps differently from amino acids.

We added intron scoring into a
multiple alignment framework, 
using a sum-of-pairs scoring policy with affine gap-scoring. 
It is NP-hard to find the alignment of two multiple alignments 
under these optimization criteria~\shortcite{aliali.NP}. 
Even the alignment of a single sequence to a multiple alignment
necessitates sophisticated techniques~\shortcite{KececiogluZhang}. 
Our solution therefore uses a gap-counting heuristic: 
namely, a gap-open penalty 
is triggered for an indel aligned with an amino acid if the 
indel is preceded by an amino acid or a phase-0 intron.  
Gap opening thus corresponds to a pattern 
$\begin{smallmatrix}\texttt{?x}\\\texttt{1-}\end{smallmatrix}$ 
or $\begin{smallmatrix}\texttt{*?x}\\\texttt{x0-}\end{smallmatrix}$.
Here, $\chone,\chzero$ are intron states for the phase-0 site, 
and $\texttt{?}$ denotes either state.
In addition, $\aaX$ denotes an arbitrary amino acid, $\aagap$ 
is the indel character, and $\aaq$ is either of the latter two.
We implemented affine gap-scoring by separate gap-open and 
-close penalties, so that gaps at the alignment extremities can be penalized 
less severely.  
Gap closing is counted for 
the patterns $\begin{smallmatrix}\texttt{-1}\\\texttt{x?}\end{smallmatrix}$ 
and $\begin{smallmatrix}\texttt{-0x}\\\texttt{x?*}\end{smallmatrix}$. 

Table~\ref{tbl:ali} gives the recurrences for a dynamic programming
algorithm that aligns an intron-annotated sequence~$S$ to 
a multiple alignment~$P$ of $h$ intron-annotated protein sequences.
In order to simplify the presentation, 
we represent the sequences in such a way that 
every odd position of~$S$ and~$P$ is a regular
residue or alignment column, annotated with 
information on the presence of phase-1 and phase-2 introns, whereas 
every even position is a phase-0 intron site.
We use $\iscore{\Sch{i}}{\Pch{j}}$
to denote the sum of intron-match scores for the intron sites associated with the 
positions~$S[i]$, $P[j]$.
We use also the shorthand $\ascore{x}{\mathbf{y}}$
to denote scoring for the alignment of an amino acid~$x$ with an amino acid profile~$\mathbf{y}$.
The algorithm uses three types of variables, 
$\varA{i}{j}$, $\varGS{i}{j}$ and $\varGP{i}{j}$,
which correspond to 
partial prefix alignments ending with 
aligned residues, gaps in~$S$, or gaps in~$P$, respectively. 
In case of~$\varGS{i}{j}$, the last indel must be aligned with an amino acid 
column, and, thus
$j$ must be odd; for~$\varGP{i}{j}$, $i$ must be odd.  

\begin{table*}
\footnotesize
\begin{align*}
\varA{i}{j} 
	& = \begin{aligned}[t]
		& \ascore{\Sch{i}}{\Pch{j}}+\iscore{\Sch{i}}{\Pch{j}}+\gape1(j)
	 	+\max\Bigl\{ 
			 \varA{i-2}{j-2}  + \iscore{\Sch{i-1}}{\Pch{j-1}} + \gapo1(j) + \gapc1(j),\\*
		&	  \varGS{i-2}{j-2} + \iscore{\Sch{i-1}}{\Pch{j-1}} + \gapo1(j) + \gapc2(j-2), 
			 \varGS{i-1}{j-2} + \iscore{\chzero}{\Pch{j-1}}   +\gapo1(j)  + \gapc2(j-2),\\*
		&	  \varGP{i-2}{j-2} + \iscore{\Sch{i-1}}{\Pch{j-1}} +\gapo2(j)  + \gapc3(j),
			 \varGP{i-2}{j-1} + \iscore{\Sch{i-1}}{\chzero^h} + \gapc2(j)
			\Bigr\}\end{aligned}
		& \text{odd $i$, odd $j$}\\
\varGS{i}{j} 
	& = \begin{aligned}[t]
		& \iscore{\chzero}{\Pch{j}} + \gape2(j)
			+ \max\Bigl\{
			  \varA{i}{j-2} + \iscore{\chzero}{\Pch{j-1}} +\gapo3(j) +\gapc1(j),\\*
			& \varGS{i}{j-2} + \iscore{\chzero}{\Pch{j-1}}, \\*
			& \varGP{i}{j-2} + \iscore{\chzero}{\Pch{j-1}} + \gapo3(j)+\gapc3(j),
			  \varGP{i}{j-1} + \iscore{\chzero}{\chzero^h} + \gapo3(j) + \gapc2(j)
			 \Bigr\}
			\end{aligned}
	& \text{odd $i$, odd $j$}\\
\varGP{i}{j} 
	& = \begin{aligned}[t]
		& \iscore{\Sch{i}}{\chzero^h} + \gape1(j)
		+ \max\Bigl\{
			  \varA{i-2}{j} + \iscore{\Sch{i-1}}{\chzero^h} + \gapo3(j),\\*
			& \varGS{i-2}{j}+ \iscore{\Sch{i-1}}{\chzero^h} + \gapo3(j) + \gapc2(j), 
			  \varGS{i-1}{j}+ \iscore{\chzero}{\chzero^h} + \gapo3(j) + \gapc2(j),\\*
			& \varGP{i-2}{j}+ \iscore{\Sch{i-1}}{\chzero^h} 
			\Bigr\}
		\end{aligned} 
	& \text{odd $i$, odd $j$}		\\
\varGS{i}{j}
	& = \begin{aligned}[t]
		& \iscore{\chzero}{\Pch{j}} + \gape2(j)
		+ \max\Bigl\{
			\varA{i-1}{j-2} + \iscore{\Sch{i}}{\Pch{j-1}} + \gapo3(j) + \gapc1(j), \\*
			& \varGS{i-1}{j-2} + \iscore{\Sch{i}}{\Pch{j-1}} + \gapo4(i,j) + \gapc4(i,j), 
			  \varGS{i}{j-2} + \iscore{\chzero}{\Pch{j-1}} , \\*
			& \varGP{i-1}{j-2} + \iscore{\Sch{i}}{\Pch{j-1}} + \gapo3(j) + \gapc3(j), 
			  \varGP{i-1}{j-1} + \iscore{\Sch{i}}{\chzero^h} + \gapo3(j) + \gapc2(j)
			\Bigr\}
		\end{aligned} 
		& \text{even $i$, odd $j$} \\
\varGP{i}{j}
	& = \begin{aligned}[t]
		& \iscore{\Sch{i}}{\chzero^h} + \gape1(j)
		+ \max\Bigl\{
			 \varA{i-2}{j-1} + \iscore{\Sch{i-1}}{\Pch{j}}  + \gapo5(j)  + \gapc5(j),\\*
			& \varGS{i-2}{j-1} + \iscore{\Sch{i-1}}{\Pch{j}} + \gapo5(j) + \gapc2(j-1),
			  \varGS{i-1}{j-1} + \iscore{\chzero}{\Pch{j}}   + \gapo5(j) + \gapc2(j-1),\\*
			& \varGP{i-2}{j-1} + \iscore{\Sch{i-1}}{\Pch{j}} + \gapo6(j) + \gapc6(j),
			  \varGP{i-2}{j} + \iscore{\Sch{i-1}}{\chzero^h}\Bigr\}
		\end{aligned} & \text{odd $i$, even $j$}
\end{align*}
\caption{Recurrences for intron-aware alignment.
	Odd positions correspond to regular amino acids and possibly phase-1 and -2 introns. 
	Even positions are placeholders for phase-0 intron sites.}\label{tbl:ali}
\end{table*}

Gaps are scored by using affine penalties, with 
gap-open, -extend, and -close 
scores, denoted by $\gapo{},\gape{},\gapc{}$.
The gap-counting heuristic implies 
that gap scores in the equations of Table~\ref{tbl:ali} 
are defined 
by the number of certain patterns in up to three consecutive 
alignment columns.
For instance, $\gapo2(j)$ equals the gap-open penalty multiplied 
by the number of such rows in~$P$ where column~$j$ contains an indel, and 
column~$j-1$ has a phase-0 intron. The corresponding pattern is described as~$\chone\aagap$. 
Table~\ref{tbl:gaps} lists the patterns for the gap-counting heuristic.

\begin{table}\small
\[
\begin{array}{llllll}
\gapo1 & \chone\aagap \text{ or } \aaX \chzero\aagap &  \gapo2 & \chone\aagap \\*
\gapo3 & \aaX & \gapo4 & \aaX \text { if $\Sch{i}$ has intron, otherwise nothing} \\*
\gapo5 & \chone \text{ or } \aaX\chzero & \gapo6 & \chone \\*
\gapc1 & \aagap\chzero\aaX \text { or } \aagap\chone\aaq & \gapc2 & \aaX \\*
\gapc3 & \chone\aaq \text{ or } \chzero\aaX &
	\gapc4 & \aaX \text{ if $\Sch{i}$ has intron, otherwise nothing} \\*
\gapc5 & \aagap\chone & \gapc6 & \chone\\*
\gape1 & \aagap & \gape2 & \aaX
\end{array}
\]
\caption{Patterns for gap counting. The index~$j$ in $\gapo{}(j)$,
$\gape{}(j)$ and~$\gapc{}(j)$
is the index for the last column  in the pattern.}\label{tbl:gaps}
\end{table}

Table~\ref{tbl:ali} does not show the
initialization of the variables, 
nor the final gap-counting: they 
employ a logic analogous to the recurrences. 
At the end of the algorithm, 
the best of~$\varA{|S|}{|P|}$, $\varGS{|S|}{|P|}$, $\varGP{|S|}{|P|}$ 
is selected, and the actual alignment is found by 
standard traceback techniques~\shortcite{Durbin}. 

We implemented the algorithm in Java. The program
iteratively realigns one sequence at a time to the 
rest of the sequences in a multiple alignment. 
Instead of sequence-dependent intron match-mismatch scores, 
the implementation uses only two parameters: one for intron conservation 
and another for intron loss/gain.

\subsection{Identification of conserved blocks}
In order to reliably identify orthologs,
we need to be able to distinguish regions of the
alignment that are highly conserved from those that are less
well-conserved. In poorly conserved regions, we cannot
confidently infer intron orthology.

\shortciteN{postprocessing} proposed post-processing 
pairwise sequence alignments 
into alternating blocks that score above a threshold parameter~$\alpha$
or below~$(-\alpha)$. 
We attained a similar 
goal by adapting 
algorithmic technniques from 
\shortciteN{segment-sets}.
The procedure separates a multiple
alignment into alternating high- and low-scoring
regions.
Using a complexity penalty~$\alpha$, 
a segmentation with~$k$ high-scoring regions 
has a segmentation score of $A-k\alpha$ where~$A$ is the sum
of scores for the aligned columns. 
Column scores are computed without gap-open and -close penalties. 
The best segmentation of an alignment of length~$\dsize$
can be found in~$O(\dsize)$ time after the column scores
are computed. 

The result of this computation is that the total of column scores in each selected 
high-scoring region will be greater than~$\alpha$.
There may be small
sub-regions of negative scores, but the
total score of such a sub-region cannot be less than~$(-\alpha)$.  
Conversely, unselected  regions score below~$(-\alpha)$
and cannot have sub-regions scoring above~$\alpha$.

\section{A likelihood framework}\label{sec:likelihood}
\subsection{Markov models of evolution}
We use a Markov model for intron evolution, as in 
previous studies~\shortcite{csuros.scenarios}. 
For the sake of generality, we describe 
the Markov model \shortcite{Steel.markov,Felsenstein} over an arbitrary
alphabet~$\alphabet$ of fixed size $\asize=|\alphabet|$.
The intron alphabet is $\alphabet=\{\chzero,\chone\}$.
A {\em phylogeny} over a 
set of species~$X$ is defined by a  rooted tree~$\tree$ 
and a probabilistic model.
The leaves are bijectively mapped to
elements of~$X$.
Each tree node~$u$ has an associated random variable~$\state(u)$, 
which is called its {\em state} or {\em label}, that takes values over~$\alphabet$. 
The tree~$\tree$ with its parameters defines the joint
distribution for the random variables~$\state(u)$.
The distribution is 
determined by the  {\em root probabilities} 
$\bigl(\pi(a)\colon a\in\alphabet\bigr)$ 
and {\em edge transition probabilities} 
$\bigl(p_e(a\rightarrow b)\colon a,b\in\alphabet\bigr)$
assigned to every edge~$e$. 
The root probabilities give the distributon of
the root state. 
Edge transition probabilities 
define the conditional distributions
$\Probcmd{\state(u_{i+1})=b}{\state(u_i)=a}=p_{u_iu_{i+1}}(a\rightarrow b)$.
Along every path away from the root, 
the node states form
a Markov chain.
The leaf states form the {\em character} $\state=(\state(u)\colon u\in X)$.
An input data set (or {\em sample}) consists 
of independent and identically distributed (iid) characters: 
$\boldsymbol{\state}=(\state_i\colon i=1,\ldots, \dsize)$.

In case of intron evolution, 
introns are generated by a two-state continuous-time Markov process with 
{\em gain} and {\em loss rates}~$\lambda_e, \mu_e\ge 0$
along each edge~$e$. 
The edge length 
is denoted by~$t_e$. 
Using standard results~\shortcite{Ross}, the transition probabilities on 
edge~$e$ with rates $\lambda_e=\lambda,\mu_e=\mu$ and length~$t_e=t$ can be written as 
$p_e(\chzero\rightarrow\chone) = \frac{\lambda}{\lambda+\mu}\bigl(1-e^{-t(\lambda+\mu)}\bigr)$,
$p_e(\chone\rightarrow\chzero) = \frac{\mu}{\lambda+\mu}\bigl(1-e^{-t(\lambda+\mu)}\bigr)$,
and $p_e(\chzero\rightarrow\chzero) = 1-p_e(\chzero\rightarrow\chone)$, 
$p_e(\chone\rightarrow\chone)=1-p_e(\chone\rightarrow\chzero)$.
In the absence of established edge lengths, we fix the edge length scaling
in such a way that $\lambda_e+\mu_e=1$.  
Independent model parameters are thus
$\pi(\chone)$, $\nu_e$ and~$t_e$ for all edges~$e$. 
It is important to allow for branch-dependent rates,
since loss and gain rates 
vary considerably between lineages
\shortcite{Jeffares.gainloss,RoyGilbert.review}.

In a maximum likelihood approach, model parameters are set by 
maximizing the likelihood of the input sample.
Let~$\mathbf{x}=(x_1,\dotsc, x_{\dsize})$ be the input data.
Every~$x_i$ is a
vector of~$\tsize$ states, 
and we write~$x_i(u)$ to denote the observed state of leaf~$u$. 
By independence, the likelihood of~$\mathbf{x}$ is the product
$\PROB\{\boldsymbol{\state}=\mathbf{x}\} = \prod_i \PROB\{\state=x_i\}$. 
Each character's likelihood 
can be computed in 
$O(\tsize)$ time,
using a dynamic programming procedure 
introduced by \shortciteN{Felsenstein.ML}.
The procedure relies on a ``pruning'' technique, 
which consists of computing the conditional likelihoods~$L_u(a)$ 
for every node~$u$ and letter~$a$. 
$L_u(a)$ is the probability 
of observing the leaf labelings in the subtree of~$u$, given 
that~$u$ is labeled with~$a$. 
%
The likelihood for the character~$x$ equals 
$L(x) = \PROB\{\state=x\}=\sum_{a\in\alphabet} \pi(a) L_{\mathsf{root}}(a)$.

Intron data are 
somewhat unusual in that 
an all-$\chzero$ character is never observed: 
the input does not include sites in which introns are absent in all of the organisms. 
The uncorrected likelihood function is therefore 
misleading, as it underestimates the probability of intron loss. 
To resolve this difficulty, 
we employ a correction technique 
proposed by 
\shortciteN{Felsenstein.restml}
for restriction sites.
(\shortciteN{csuros.scenarios} describes an alternative technique based on expectation maximization.)
We compute the likelihood under the condition that 
the input does not include all-$\chzero$ characters. 
We use therefore the 
corrected likelihood $L'(x) = \frac{L(x)}{1-L(\chzero^\tsize)}$, 
and maximize $L'=\prod_i L'(x_i)$. 

\subsection{Ancestral events in intron evolution}\label{ss:ancestral}
Our goal is to give a reliable characterization of the dynamics of intron evolution. 
In particular, we aim to give estimates for intron density in ancestral species, 
and for intron loss and gain events on the edges. 
Notice that the estimation method needs to 
account for ancestral introns even if they got eliminated in all 
descendant lineages.
It is possible to do that with the help of conditional expectations, 
which fit naturally into a
likelihood framework. 

For an observed character~$x$,
we define 
{\em upper conditional likelihoods}~$U_u(a)$ so that $U_u(a)L_u(a)=\PROB\{\xi=x, \xi(u)=a\}$.
Upper conditional likelihoods are computed 
with dynamic programming, from the root towards the leaves, 
in~$O(\tsize\dsize)$ time~\shortcite{csuros.scenarios},
even for non-binary trees.
Similar computations are routinely used in 
DNA and protein likelihood 
maximization programs~\shortcite{MOLPHY,PhyML}. 
Here we allow irreversible probabilistic models, which 
explains why~$U_u(a)$ must be computed in a top-down fashion
in~\eqref{eq:upper}.

The posterior probability for the state at node~$u$ is 
\[
q_a^{(u)}(x) = \Probcmd{\state(u)=a}{\state=x} = \frac{U_u(a) L_u(a)}{\sum_{b\in\alphabet} U_u(b)L_u(b)}.
\]
The posterior probabilities for state changes on an edge~$uv$ are 
\begin{multline*}
q_{ab}^{(v)}(x)= \Probcmd{\state(u)=a,\state(v)=b}{\state=x}\\*
	= U_u(a)L_u(a) \frac{p_{uv}(a\rightarrow b)L_v(b)}{\sum_{a'} p_{uv}(a\rightarrow a')L_v(a')}. 
\end{multline*}

Now, the number of ancestral introns is estimated as 
the conditional expectation
$N_u = \dsize_0 q_{\chone}^{(u)}(\chzero^\tsize)+\sum_i q_{\chone}^{(u)}(x_i)$.
The formula takes into consideration
unobserved intron sites, by estimating 
their number $\dsize_0=\dsize \frac{L(\chzero^\tsize)}{1-L(\chzero^\tsize)}$
as the mean of a negative binomial random variable. 
The number of intron state changes is 
estimated as $N_v(a\rightarrow b) = \dsize_0 q_{ab}^{(v)} (\chzero^\tsize)+\sum_i q_{ab}^{(v)}(x_i)$.
In particular, $N_v(\chone\rightarrow\chzero)$ is the number of introns lost, and 
$N_v(\chzero\rightarrow \chone)$ is the number of introns gained 
on the edge leading to~$v$. 

In order to compute~$U$, we initialize $U_{\mathsf{root}}(a) = \pi(a)$.
On every edge~$uv$, 
\begin{multline}\label{eq:upper}
U_v(b) \\
	= \sum_{a\in\alphabet} U_u(a)p_{uv}(a\rightarrow  b) \prod_{w\in\mathsf{siblings}(v) }
	\sum_{a'\in\alphabet}p_{uw}(a\rightarrow a')L_w(a')\\
	= \sum_{a\in\alphabet} U_u(a) L_u(a)
		\frac{p_{uv}(a\rightarrow  b) L_v(b)}{\sum_{a'\in\alphabet} p_{uv}(a\rightarrow a') L_v(a')}.
\end{multline}

\section{Rapid computation of the likelihood}\label{sec:compress}
There are many heuristics that accelerate likelihood-based 
phylogenetic reconstruction~\shortcite{SEMPHY,PhyML},
which mostly concentrate on the exploration of the tree space. 
We propose an improvement to the evaluation of 
the likelihood function, 
which normally takes linear time
in the input size. 
Our evaluation method yields an~$O(\log\dsize)$ speedup for 
typical trees.
We use it to optimize the parameters of intron evolution
on a known tree, but the method is generally applicable to 
phylogenetic likelihood problems where 
the likelihood is numerically optimized.
The key idea is that it is enough 
to carry out the pruning algorithm once 
for every different labeling within a subtree. 
Different labelings within each subtree can be computed 
in a preprocessing step. 
Subsequent evaluations of the likelihood function with different model parameters 
are faster, and depend on the number of different labelings in the data set. 

Here we describe how the preprocessing step can be carried out in~$O(\tsize\dsize)$
time. 
Secondly, we analyze the computational complexity of subsequent evaluations, and show
that an~$O(\log\dsize)$ speedup is 
achieved for almost all random trees in the Yule-Harding model.
The latter analysis 
produces some novel concentration results on the number of subtrees with 
a fixed size~$k$. 

To our knowledge, the closest idea 
to ours
was articulated by \shortciteN{AxML}. 
Specifically, 
they proposed identifying 
characters in which leaves in a 
subtree have identical labels. 
They reported that in benchmark experiments
with nucleotide sequences, 
likelihood optimization was accelerated by 12--15\%
through this technique. 
The technique relies entirely on the fact that 
alignments of closely related sequences 
exhibit high levels of identity, and cannot be
extended to non-identical labelings. 

\subsection{Yule-Harding model}
The Yule-Harding distribution is encountered in random birth and 
death models of species and in coalescent models~\shortcite{Felsenstein},
and is thus one of the most adequate random models for phylogenies. 
In one of the equivalent formulations, a random tree
is grown by adding leaves one by one in a random order.
The leaves are first 
numbered by
using a random uniform permutation
of the integers~$1,2,\dotsc,\tsize$.
Leaves are joined to the tree in an iterative procedure. 
In step 1, the tree is just leaf~1 on its own. In step 2, 
the tree is a ``cherry'' with leaves~$1$ and~$2$. 
In each subsequent 
step~$i=3,\dotsc,\tsize$, a random leaf~$Y_i$ is picked uniformly from the set 
$\{1,2,\dotsc,i-1\}$. The new leaf~$i$ is added 
to the tree as the sibling of~$Y_i$, forming a cherry: 
a new node is placed on the edge leading to~$Y_i$ and~$i$ 
is connected to it. 
The resulting random tree in iteration~$\tsize$ 
has the Yule-Harding distribution~\shortcite{Harding}. 

\begin{figure}
\begin{fmpage}{0.98\columnwidth}
\footnotesize
\begin{algolist}{C}
\item[] Algorithm \Algo{Compress}
\item\mt\ \textbf{for} every leaf $u$ \textbf{do}
\item\mtt\ initialize $a[1..\asize]\setto 0$ // {\em $a[j]$ is the first occurrence of state $j$}
\item\mtt\ \textbf{for} $i\setto 1,\dotsc,\dsize$ \textbf{do}
\item\mttt\ \textbf{if} $a[x_i(u)] = 0$ \textbf{then} $a[x_i(u)]\setto i$
\item\mttt\ set $h_i(u)\setto a[x_i(u)]$  
\item\mt\ \textbf{for} every non-leaf node $u$ in a post-order traversal \textbf{do}
\item\mtt\ initialize the map $H\colon \{1,\dotsc,\dsize\}^2\mapsto \{1,\dotsc,\dsize\}$ as empty \label{line:compression.map.init}
\item\mtt\ let $v,w$ be the left and right children of $u$
\item\mtt\ \textbf{for} $i\setto1,\dotsc,\dsize$ \textbf{do}
\item\mttt\ \textbf{if} $H(h_i(v),h_i(w)) = \mathsf{null}$ \textbf{then} $H(h_i(v),h_i(w))\setto i$ \label{line:compression.map.set}
\item\mttt\ set $h_i(u)\setto H(h_i(v),h_i(w))$ \label{line:compression.map.get}
\end{algolist}
\end{fmpage}
\caption{Computing the auxiliary arrays from which
	the multiplicities~$\multiplicity_u$ are obtained.
	}\label{fig:compression}
\end{figure}

\subsection{Preprocessing}\label{ss:compression}
The likelihood can be computed faster 
by first identifying 
the different~$x_i$ values, along with their
multiplicity in the data. 
For large trees, the input typically consists of many different labelings, 
but for small trees with~$\tsize<\log_{\asize}\dsize)$, 
the compression is useful, since the number of different~$x_i$
values is bounded by~$\asize^{\tsize}<\dsize$. 
In order to exploit the benefits of compression, 
we extend it to every subtree. 

\begin{definition}\label{def:compression}
Define the multiset of observed labelings 
for every node~$u$ as follows. 
Let~$\tsize'$ denote the 
number of leaves in the subtree rooted at~$u$, 
and let~$u_1,\dotsc,u_{\tsize'}$ denote those leaves. 
Define~$\multiplicity_u(y)$ for every 
labeling~$y\in\alphabet^{\tsize'}$ of 
the leaves in the subtree of~$u$ 
as the number of times~$y$ is observed in the data:
\[
\multiplicity_u(y) = \sum_{i=1}^{\dsize} \chi\Bigl\{  \forall\, k\colon y_k = x_{i}(u_k)\Bigr\},
\] 
where~$\chi\{\cdot\}$ denotes the indicator function that takes the value 1 if its argument 
is true, otherwise it is~0. 
Define also the set of observed labelings
\[
\oszlopok_u = \Bigl\{ y\in\alphabet^{\tsize'}\colon \multiplicity_u(y) >0  \Bigr\}.
\]
\end{definition}

The multisets of observed labelings 
are computed in the preprocessing step. 
The likelihood function is evaluated 
subsequently by computing the conditional likelihoods 
at each node~$u$ for 
the labelings of~$\oszlopok_u$ only, in~$O(|\oszlopok_u|)$ 
time.

In order to compute~$\multiplicity_u$, we use a 
recursive procedure. 
It is important to avoid 
working with the $O(\tsize)$-dimensional vectors~$y$ 
of Definition~\ref{def:compression} directly, otherwise
the preprocessing may take superlinear time in~$\tsize$. 
For that reason, every labeling $y\in\oszlopok_u$ 
is represented  
by the index~$i$ for which~$x_i$ 
is the first occurrence of~$y$. 
Accordingly, we compute an auxiliary array~$h_i(u)$,
which stores the first occurrence of each labeling~$x_i$ 
in~$u$'s subtree.
In particular, $h_i(u)=i'$ if~$i'$ is the smallest index such that 
	$x_{i}(u_k)=x_{i'}(u_k)$ for all~$k$ where~$u_k$ 
	are the leaves in~$u$'s subtree.
Figure~\ref{fig:compression}
shows that the values~$h_i(u)$ can be computed in a post-order traversal.
After~$h_i(u)$ are computed for all~$i$ and~$u$, 
the multiplicities~$\multiplicity_u$ and observed labelings~$\oszlopok_u$ 
are straightforward to calculate in linear time. 
The map~$H$ in Lines~\ref{line:compression.map.init}--\ref{line:compression.map.get}
is sparse with at most~$\dsize$ entries, and can be implemented as 
a hash table so that accessing and updating it takes~$O(1)$ time. 
Consequently, Algorithm~\Algo{Compress} takes $O(\tsize\dsize)$ time. 

\begin{figure}
\begin{fmpage}{0.98\columnwidth}
\footnotesize
\begin{algolist}{L}
\item[] Algorithm \Algo{LogLikelihood}
\item\mt\ \textbf{for} every leaf $u$ do
\item\mtt\ set $L_u[a][a']\setto \chi\{a=a'\}$ for all $a\in\oszlopok_u$ and $a'\in\mathcal{\alphabet}$
\item\mt\ \textbf{for} every non-leaf node $u$ in a post-order traversal \textbf{do}
\item\mtt\ \textbf{for} every labeling $y\in\oszlopok_u$ \textbf{do} 
\item\mttt\ let $u_1,u_2$ be the children of $u$, and let $y_1,y_2$ be their subtree labelings
\item\mttt\ \textbf{for} every $a\in\alphabet$
\item\mtttt\ $L_u[y][a] \setto \prod_{j=1,2}\sum_{a'\in\alphabet} p_{uu_j}(a\rightarrow a') L_{u_j}[y_j][a']$
\item\mt\ set $\mathsf{logL}\setto 0$
\item\mt\ \textbf{for} every labeling $y\in\oszlopok_{\mathsf{root}}$ \textbf{do}
\item\mtt\ $\mathsf{logL} \setto \mathsf{logL} +\multiplicity_{\mathsf{root}}(y) \cdot \sum_{a\in\alphabet} \log \bigl(\pi(a) L_{\mathsf{root}}[y][a]\bigr)$
\item\mt\ \textbf{return} $\mathsf{logL}$
\end{algolist}
\end{fmpage}
\caption{Computing the log-likelihood using the observed labelings.
	}\label{fig:likelihood}
\end{figure}

\subsection{Evaluating the likelihood function}
After the preprocessing step, 
the conditional likelihoods
are computed only for the different labelings within each 
subtree. 
Figure~\ref{fig:likelihood} shows the evaluation of the likelihood function.
The 
running time for the algorithm
is $O(\nlabelings)$ where~$\nlabelings$ is the total number of different 
labelings within all subtrees:
\[
\nlabelings = \sum_u |\oszlopok_u|. 
\]
If~$\tsize_u$ denotes the number of leaves in the subtree rooted 
at~$u$, then~$\oszlopok_u$ has at most~$\asize^{\tsize_u}$ 
elements. Hence,~$\nlabelings$ is bounded as
\begin{equation}\label{eq:nlabelings.bound}
\nlabelings \le \sum_u\min\{\asize^{\tsize_u}, \dsize\}.
\end{equation}
Observe that by sheer number of arithmetic operations,
it is always worth evaluating the likelihood function
this way. The worst situation is that of 
a caterpillar tree (where every inner node has a leaf child).
In that case, there are only a few ($\lfloor\log_{\asize} \dsize\rfloor-1$) 
non-leaf nodes for which we can compress the data, 
and it is possible to construct an artificial data set 
in which there are~$\dsize$ different 
labelings for $\tsize-O(\log \dsize)$ subtrees. 
Caterpillar trees, however, are rare in 
phylogenetic analysis. Typical phylogenies
have fairly balanced subtrees~\shortcite{Aldous.beta}. 

In what follows, we examine the bound of~\eqref{eq:nlabelings.bound} 
more closely in the Yule-Harding model. 
The analysis relies on a characterization of 
the random number of subtrees with 
a given size, as expressed in 
Theorems~\ref{tm:sizes.mean}
and~\ref{tm:sizes.concentration}.

\begin{theorem}\label{tm:sizes.mean}
Consider random evolutionary trees with~$\tsize$ leaves in 
the Yule-Harding model. 
Let~$\sizecount{k}$ denote the 
number of subtrees with~$k=1,\dotsc,\tsize-1$ leaves in 
a random tree.  
The expected value of~$\sizecount{k}$ is 
\begin{equation}\label{eq:sizes.mean}
\EXP \sizecount{k} = 2\tsize\Bigl(\frac 1k-\frac1{k+1}\Bigr).
\end{equation}
Trivially, $\EXP \sizecount{\tsize}=1$.
\end{theorem}
\begin{proof}
\shortciteN{Heard.subtrees} derives Equation~\eqref{eq:sizes.mean}
by appealing to a P\'olya urn model.
An equivalent result is stated by \shortciteN{Devroye.subtrees.normal}.
\QED\end{proof}
%

\begin{theorem}\label{tm:sizes.concentration}
For all $\epsilon>0$, 
\begin{subequations}\label{eq:sizes.tail}
\begin{align}
\PROB\bigl\{\sizecount{k} \le \EXP\sizecount{k}-\epsilon\bigr\}
	& \le e^{-\frac{\epsilon^2}{2\tsize}}; \label{eq:sizes.tail.left}\\*
\PROB\bigl\{\sizecount{k} \ge \EXP\sizecount{k}+\epsilon\bigr\}
	& \le e^{-\frac{\epsilon^2}{2\tsize}}.
\end{align}
\end{subequations}
\end{theorem}
\begin{proof}
Consider the random construction of the tree. 
Let~$Y_i$ denote the random leaf picked in step~$i$ to which leaf~$i$ gets connected, 
for $i=3,4,\dotsc,\tsize$.  
Each random variable~$Y_i$ is uniformly distributed over the set~$\{1,2,\dotsc,i-1\}$, and 
$Y_3, Y_4, \dotsc, Y_{\tsize}$ are independent.
Moreover,  
they completely determine 
the tree~$T$ at the end of the procedure. 
Consequently, $\sizecount{k}$~is a function of~$(Y_i\colon i=3,\dotsc, \tsize)$:
$\sizecount{k} = f(Y_3,Y_4,\dotsc, Y_{\tsize})$. 
The key observation for the concentration result is that 
if we change the value of only one of the~$Y_i$ in the series, then  
$\sizecount{k}$ changes by at most two.
In order to see this, consider what happens to the tree~$T$, 
if we change the value of exactly one of the~$Y_i$ from~$y$ to~$y'$. 
Such a change corresponds to a ``subtree prune and regraft'' 
transformation~\shortcite{Felsenstein}. Specifically, the subtree~$T_i$, defined 
as the child tree of the lowest common ancestor~$u$ of~$y$ 
and~$i$ containing the leaf~$i$, is cut from~$T$, and is reattached to one of the
edges on the path from the root to~$y'$. 
Now, such a transformation does not affect~$\sizecount{k}$ by much. 
Notice that subtree sizes are strictly monotone decreasing from the root 
on every path. 
On the path from the root to~$u$, subtree sizes decrease by 
the size~$\tau$ of~$T_i$, and~$u$ disappears, contributing a change of $+1$, $0$ or~$(-1)$ 
to~$\sizecount{k}$. 
(At most one subtree of size~$\tau+k$ that contains $T_i$ now has size~$k$,
	and at most one subtree of original size~$k$ is not counted anymore: it may be $u$'s subtree
	itself, 
	or a subtree above it.)
An analogous argument shows that regrafting contributes a change of~$+1$, $0$, 
or~$(-1)$. Hence, 
the function~$f(\cdot)$ is such that by changing one of its 
arguments, its value changes by at most~$2$. 
As a consequence, McDiarmid's 
inequality~(\citeyear{McDiarmid}) can be applied
to bound the probabilities of large deviations for~$\sizecount{k}$.
In particular, for all~$\epsilon>0$, 
\[
\PROB\bigl\{f(Y_3,\dotsc,Y_{\tsize})-\EXP f(Y_3,\dotsc,Y_{\tsize})\le -\epsilon\bigr\}
	\le e^{-2\epsilon^2/c^2}
\]
where~$c^2=\sum_{i=3}^{\tsize} c_i^2$ and 
\begin{multline*}
c_i = \max_{y_3,\dotsc,y_{\tsize},y,y'}\Bigl|f(y_3,\dotsc, y_{i-1}, y, y_{i+1},\dotsc, y_{\tsize}) \\*
	- f(y_3,\dotsc, y_{i-1}, y', y_{i+1},\dotsc, y_{\tsize})\Bigr|. 
\end{multline*}
Since $c_i\le 2$ for all~$i$, 
$
\PROB\bigl\{\sizecount{k} \le \EXP\sizecount{k}-\epsilon\bigr\}
	\le e^{-\frac{\epsilon^2}{2(\tsize-2)}}$,
implying Eq.~\eqref{eq:sizes.tail.left}.
An identical bound holds for the right-hand tail of the distribution.
\QED\end{proof}

\textsc{Remark.}\ \  
The particular case of~$k=2$
was considered by~\shortciteN{McKenzieSteel.cherries}. 
They showed that the distribution of~$\sizecount{2}$ 
is asymptotically normal with mean~$\tsize/3$ and 
variance~$2\tsize/45$. 
The result suggests that the best 
constant factor in the exponent of Eqs.~\eqref{eq:sizes.tail}
is~$45/8$ for~$i=2$, instead of~$\frac12$.
\shortciteN{Rosenberg.subtrees} gave exact formulas for the variance of~$\sizecount{k}$. 
He showed that the variance of~$\sizecount{k}$ is $(2+o(1))\frac{\tsize}{k^2}$. 
The variance formulas were given earlier in a different context by \shortciteN{Devroye.subtrees.normal}.
(See also the discussion by \shortciteN{BlumFrancois.Sackin}.)
The result suggests that by analogy with the cherries, the best 
constant factor in the exponent is~$(1/4+o(1))k^2$.
It is thus plausible that the probability is properly bounded 
by $1-o\bigl(\tsize \log^{-2}\dsize\bigr)$
in Theorem~\ref{tm:compute} below. 

\begin{theorem}\label{tm:compute}
With probability~$1-o\Bigl(\frac{\tsize}{\log^4\dsize}\Bigr)$,
the likelihood function can be evaluated for random
trees in the Yule-Harding model in
$O\Bigl(\frac{\tsize\dsize}{\log \dsize}\Bigr)$
time after an initial 
preprocessing step that takes~$O(\tsize\dsize)$ time. 
Evaluating the likelihood function takes~$O(\tsize\dsize)$
time in the worst case,
and $O(\tsize\dsize\log^{-1} \dsize)$ time on average. 
\end{theorem}

We need the following lemma for the proof of Theorem~\ref{tm:compute}. 
\begin{lemma}\label{lm:sum.q^i/i}
For all $t\ge 4$, $\sum_{k=1}^t \frac{2^k}{k(k+1)}<\frac{2^t}{t+1}$. 
For all $t\ge 1$ and $\asize=3,4,\dotsc$, 
$\sum_{k=1}^t \frac{\asize^k}{k(k+1)}\le\frac{\asize^t}{t+1}$. 
\end{lemma}
\begin{proof}
The proof is straightforward by induction in~$t$. 
Notice that the right order of magnitude is 
$\sum_{k=1}^t \frac{\asize^k}{k(k+1)} = \Theta\bigl(t^{-2}\asize^t\bigr)$
but we need a bound for all~$t$. 
\QED\end{proof}

\begin{table}\footnotesize
\begin{center}
\begin{tabular}{rrcrrrr}
$\tsize$ & $\dsize$ & $\asize$ & $\tsize\dsize$ & $\tsize|\oszlopok_{\mathsf{root}}|$ & bound & $\nlabelings$ \\*
\hline
8 & 7236 & 2 & 101304 & 1386 & 368 & 183 \\*
18 & 8044 & 2 & 273496 & 19142 & 16764 & 1196\\*
47 & 5216 & 3 & 479872 & 309120 & 65743 & 10305 \\
\hline
\end{tabular}
\end{center}
\caption{Effect of the compression on three data sets. 
	The fourth column quantifies the 
	direct evaluation method, 
	the fifth column 
	quantifies the effect of the compression
	restricted to the root, 
	the sixth column corresponds to the bound
	of Eq.~\eqref{eq:nlabelings.bound}
	and the seventh column gives the exact value of~$\nlabelings$.  
	The first data set is from Rogozin et al. (2003), 
	the second data set is the one analyzed here in~\S\ref{ss:app.ancestors},
	and the third one is an unpublished data set 
	we have worked on, where an ambiguity character is included in the alphabet. 
	}\label{tbl:practice}
\end{table}

\begin{proof}[Proof of Theorem~\ref{tm:compute}]
The preprocessing~(Fig.~\ref{fig:compression}) takes $O(\tsize\dsize)$ 
time as discussed in~\S\ref{ss:compression}.
The evaluation of the likelihood function (Fig.~\ref{fig:likelihood}) takes 
$O(\nlabelings)$ time. By Eq.~\eqref{eq:nlabelings.bound},
\begin{equation}\label{eq:likelihood.time}
\nlabelings \le \sum_{k=1}^{\tsize} 
	\sizecount{k} \min\{\asize^k, \dsize\}
	= \sum_{k=1}^{\lfloor \log_{\asize} \dsize\rfloor} \sizecount{k}\asize^k
	+ \sum_{k=1+\lfloor \log_{\asize} \dsize\rfloor}^\tsize \sizecount{k}\dsize.
\end{equation}
Let $t = \lfloor \log_{\asize} \dsize\rfloor$.
By Theorem~\ref{tm:sizes.mean} and Lemma~\ref{lm:sum.q^i/i}, if~$\dsize\ge 16$ or $\asize\ge 3$,
\begin{align}
\EXP \nlabelings 
	& \le 2\tsize
		\sum_{k=1}^t \frac{\asize^k}{k(k+1)} 
		+ \dsize\Bigl(\frac{2\tsize}{t+1}-1\Bigr)\nonumber\\*
	& \le 2\tsize \frac{\asize^t}{t+1} + 2\tsize\frac{\dsize}{t+1}
		 \le \frac{4\tsize\dsize}{t+1}, \label{eq:nlabelings.mean}
\end{align}
which proves our claim about the average running time. 
Now, let $\epsilon=\frac{n}{2t(t+1)}$.
Plugging~$\epsilon$ into Theorem~\ref{tm:sizes.concentration},
we get that 
\begin{equation}\label{eq:epsilon.bound}
\PROB\Bigl\{\bigl|\sizecount{k}-\EXP\sizecount{k}\bigr| \ge\epsilon \Bigr\}
	\le 2 \exp\Biggl(-\frac{\tsize}{8}\Bigl(\frac1t-\frac{1}{t+1}\Bigr)^2\Biggr). 
\end{equation}
Let~$\mathcal{E}_k$ denote the event that 
$\bigl|\sizecount{k}-\EXP{\sizecount{k}}\bigr| < \epsilon$
for $k=1,\dotsc, t$, and let 
$\mathcal{E}_{t+1}$ denote the event that 
$\bigl|\sum_{k=1+t}^{\tsize}\sizecount{k}-\EXP\sum_{k=1+t}^{\tsize}\sizecount{k}\bigr|<t\epsilon$.
Since
$\sum_k \sizecount{k} = 2\tsize-1$,
$\cap_{k=1}^t\mathcal{E}_k$ implies~$\mathcal{E}_{t+1}$. 
By~\eqref{eq:epsilon.bound},
\[
\PROB \bigcap_{k=1}^{t+1} \mathcal{E}_k
	\ge 1-\sum_{k=1}^t \PROB\bar{\mathcal{E}}_k 
	\ge 1-2 t \exp\Biggl(-\frac{\tsize}{8}\Bigl(\frac1t-\frac{1}{t+1}\Bigr)^2\Biggr),
\]
where~$\bar{\mathcal{E}}_k$ denotes the complementary event to~$\mathcal{E}_k$. 
Now, $\cap_{k=1}^t\mathcal{E}_k$ also implies that 
$\nlabelings\le \frac{5\tsize\dsize}{t+1}$.
Since the likelihood computation takes~$O(\nlabelings)$ time, 
the theorem holds. 
\QED\end{proof}

Theorem~\ref{tm:compute}
underestimates the actual 
speedups that the compression method brings about. Table~\ref{tbl:practice}
shows that compression
results in a 50--500 fold speedup 
of the likelihood evaluation in 
practice. 
Notice also that 
the constants hidden behind the asymptotic notation 
are quite different between the 
preprocessing and evaluation 
steps:
costly floating point 
operations are avoided in the 
preprocessing step. 
It is important to stress that 
the theoretical analysis does not 
rely on similarities in the input data:
Theorem~\ref{tm:compute} and 
the bound of~\eqref{eq:nlabelings.bound} 
hold for any sample of size~$\dsize$.
Real-life data are expected to behave even better, as 
Table~\ref{tbl:practice} illustrates.

Even though the theorem holds for 
arbitrary alphabets, the compression is less effective 
for large alphabets (amino acids for example)
in practice. For DNA sequences, however, it should still be valuable:
we conjecture that compression would accelerate likelihood 
optimization by at least an order of magnitude.

We implemented Algorithms~\Algo{Compress} and \Algo{LogLikelihood}, 
along with likelihood maximization and 
the posterior calculations of \S\ref{ss:ancestral}
in a Java package. 
Likelihood is maximized by 
setting loss and gain parameters 
for the edges, by using mostly 
the Broyden-Fletcher-Goldfarb-Shanno
method~\shortcite{NR} whenever 
possible, and occasionally
Brent's line minimization \shortcite{NR}
for each parameter separately. 

\section{Applications}\label{sec:applications}
\subsection{Ancient paralogs}
In our first example, intron-aware alignment 
was used to reject a hypothesis 
about whether lack of intron sharing between 
homologous genes is due to 
poor protein alignments. 

We used intron-aware alignment in a 
study about ancient eukaryotic paralogs~\shortcite{Sverdlov.ancientdup}. 
In the study, 157 homologous gene families 
were examined across six species ({\em A.~thaliana}, {\em H.~sapiens}, 
	{\em C.~elegans}, {\em D.~melanogaster}, {\em S.~cerevis\ae} 
	and {\em S.~pombe}). 
These families were notable because they contained 
paralogous members in multiple eukaryotic species, but not 
in prokaryotes, and, thus, presumably underwent 
duplication in the lineage leading to LECA. 
Ancient paralogs within and across 
species share very few (in the order of a few percentages) introns.
The finding is quite surprising, as 
recent paralogs, resulting from lineage-specific duplications, 
and also orthologs between human and Arabidopsis 
agree much more in their intron sites~\shortcite{Rogozin.data}. 
Consequently, 
ancient paralogs either lacked introns at the time of their duplication,
or their duplication involved removal of introns.  

In one of the data validation steps for the study, we 
used intron-aware alignment
with very high intron match rewards.
With larger rewards for intron matches, 
more introns line up in the alignment, but
the protein alignment gets worse. 
Even by 
corrupting the protein alignment, we were not able to 
achieve intron sharing levels 
similar to that of human-Arabidopsis orthologs.
The lack of intron agreement is therefore not 
an artifact of the protein alignments. 

More details of our study will be described in a 
forthcoming publication~\shortcite{Sverdlov.ancientdup}.

\subsection{Intron-rich ancestors}\label{ss:app.ancestors}
We compiled a data set with 18 eukaryotic species to 
give a comprehensive picture of spliceosomal evolution 
among Eukaryotes. 

\begin{table}[h]\footnotesize
\begin{tabular}{llll}
Species & Abbreviation & Assembly & Source \\*
\hline
{\em Homo sapiens}				& Hsap & 36.2 & R \\*
{\em Rattus norvegicus}			& Rnor & RGSC 3.4 & E \\*
{\em Takifugu rubripes}			& Trub & FUGU 4.0 & E \\*
{\em Danio rerio}				& Drer & Zv6 & E \\*
{\em Drosophila melanogaster}	& Dmel & BDGP 4.3 & R\\*
{\em Anopheles gambi{\ae}}		& Agam & AgamP3 & R \\*
{\em Apis mellifera}			& Amel & AMEL4.0 & R \\*
{\em C{\ae}norhabditis elegans}	& Cele & WS160 & R \\*
{\em C{\ae}norhabditis briggs{\ae}} & Cbri & CB25 & W \\*
{\em Saccharomyces cerevisi{\ae}} & Scer & 2.1 & R \\*
{\em Neurospora crassa} OR74A	& Ncer	& & R \\*
{\em Schizosaccharomyces pombe} 972h- & Spom & 1.1 & R \\*
{\em Ustilago maydis} 521		& Umay & & R \\*
{\em Cryptococcus neoformans v. n.} JEC21 & Cneo & 1.1 & R\\*
{\em Oryza sativa ssp. japonica} & Osat & RAP 3 & R\\*
{\em Arabidopsis thaliana} & Atha & 6.0 & R\\*
{\em Plasmodium falciparum} 3D7 & Pfal & 1.1 & R \\*
{\em Plasmodium berghei} str. ANKA & Pber & & R
\end{tabular}
\caption{Data sources and species abbreviations. 
	R: RefSeq release 20, E: Ensemble release 42, W: WormBase release 160.}\label{tbl:data}
\end{table}

\subsubsection*{Data preparation}
Genbank flatfiles and protein sequences
were downloaded from 
RefSeq \shortcite{RefSeq2007}
and
Ensembl \shortcite{Ensembl2007}. 
Exon-intron annotation was extracted from the 
Genbank flatfiles.
The {\em C.~briggs{\ae}} protein sequences and genome annotation were downloaded from WormBase
\shortcite{WormBase2007}, and intron annotation was extracted from 
the GFF annotation file. 
Table~\ref{tbl:data} lists the data sources and the species 
abbreviations. 
We used the 684 ortholog sets, each corresponding to a 
cluster of orthologous groups, or KOG \shortcite{COG.new},
from the study 
of \shortciteN{Rogozin.data} as ``seeds'' for 
compiling a set of putative orthologs for our species. 
For each seed (consisting of homologous protein sequences 
for eight species), we performed a position-specific iterated 
BLAST search \shortcite{BLAST}. In case of Plasmodium species, 
we used three iterations against a database of all protozoan 
peptide sequences in RefSeq. We used an E-value cutoff of~$10^{-9}$ 
for retaining candidates. Each candidate hit was then 
used as a query in a reversed position-specific BLAST (rpsblast)
search \shortcite{CDD2007} against the KOG database. 
Candidates were retained after this point if they had 
the highest scoring hit (by rpsblast) against the same 
KOG as the KOG of the seed data, and they scored within~80\% 
of the best such hit for the species.

\begin{figure*}
\centerline{\includegraphics[width=0.4\textwidth]{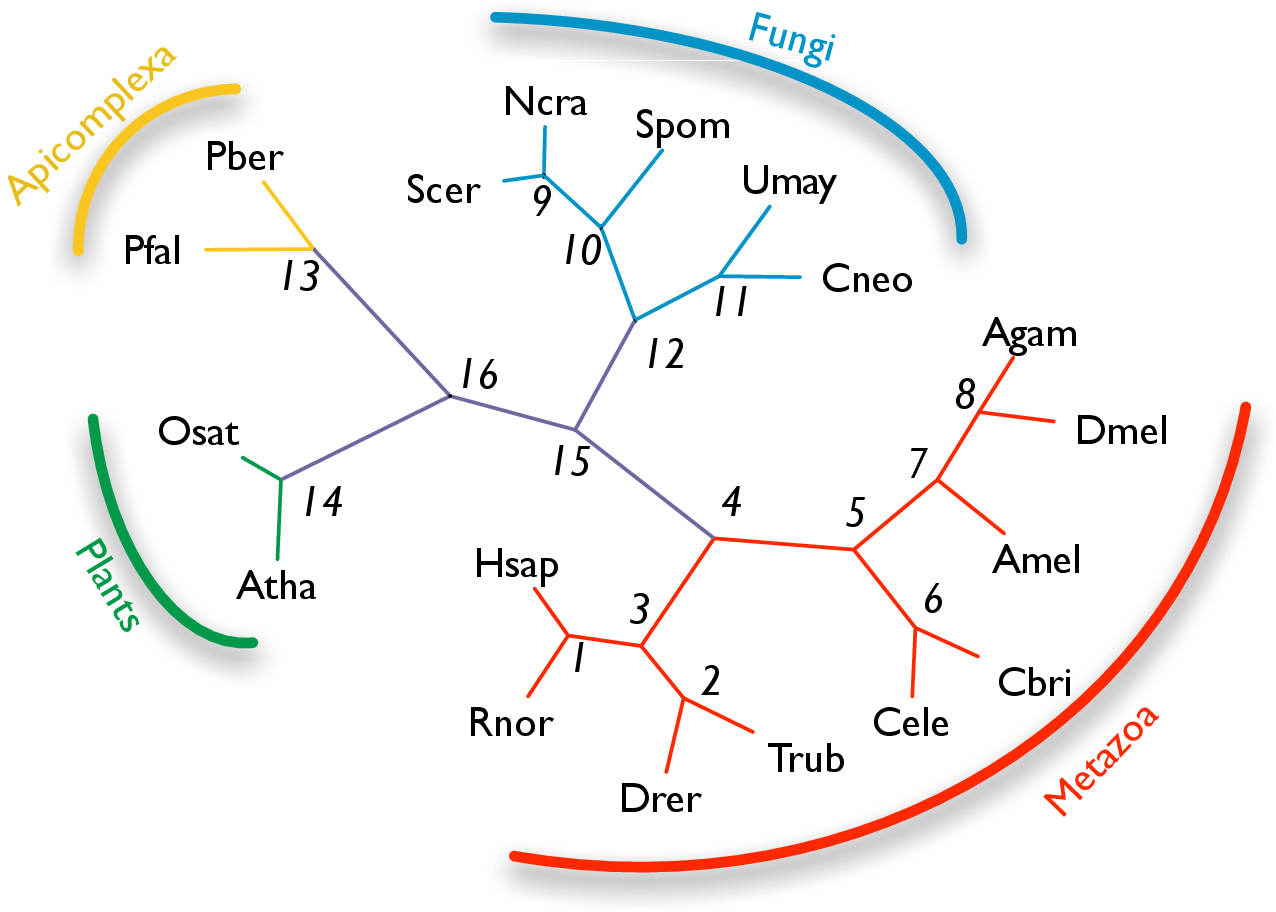}\hfill
	\includegraphics[width=0.6\textwidth]{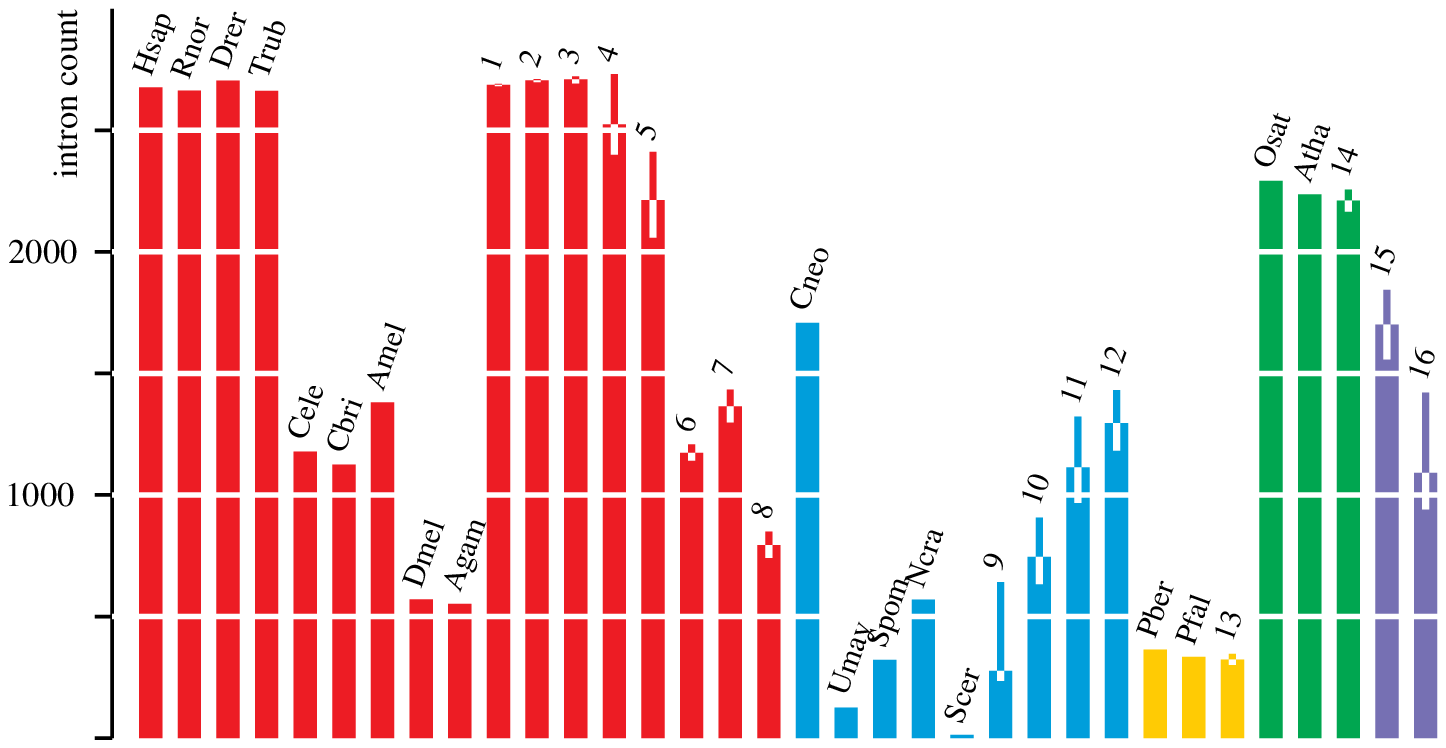}}
\caption{Phylogeny for the 18 species and 
	estimated intron counts at ancestors.
	Ancestral nodes are identified by the numbers 1--16.
	Error bars denote 95\% confidence intervals computed from 
	1000 simulated data sets.
	We rooted the tree at node 16 (Bikonts) for computational purposes.
	}\label{fig:posteriors}
\end{figure*}

From the resulting set of paralogs, we selected 
a putative ortholog set in the following manner. 
For each KOG, we selected all possible triples 
of human-Arabidopsis-Saccharomyces paralogs 
and kept the triple that had the highest 
score in alignments built by MUSCLE~\shortcite{MUSCLE}.
Alignment score was computed using the VTML240 matrix~\shortcite{VTML}. 
Additional putative orthologs were added for one species at a time, 
by aligning each paralog individually to the current 
profile, and keeping the one that gave the largest 
alignment score. At this iterative addition, scoring was done 
with the VTML240 matrix, by summing 
the five highest pairwise scores between a candidate and already included 
sequences. 

The resulting sets were then realigned using MUSCLE, and then realigned again 
using our intron-aware alignment with a gap penalty of~300, gap-extend penalty of~11,
VTML240 amino acid scoring, intron-match scores of~300 and intron-mismatch penalties of~20.
(These latter were established using different intron-match and -mismatch scores,
and selecting the ones that gave the fewest number of intron sites while decreasing 
the score of the 
implied protein alignment
by less than~0.1\%.)

Conserved portions were extracted using our segmentation program with 
a complexity penalty of~$\alpha=400$ 
(larger values gave identical segmentation results, 
and lower 
values resulted in too many scattered blocks). We penalized indels 
with an infinitely large value to exclude gap columns. 
Phase-0 introns falling on the boundaries of conserved blocks were excluded. 
Intron presence and absence in the aligned data 
was then extracted to produce the data for the likelihood programs. 

\subsubsection*{Results}
Figure~\ref{fig:posteriors}
shows the estimated intron densities for ancestral species. 
It is notable that ancestral nodes such as
the bilaterian ancestor (node 4), the ecdysozoan ancestor (node 5), 
the opisthokont ancestor (node 15), 
and the bikont ancestor (node 16) 
all have very high intron densities, 
surpassing most previous estimates~\shortcite{Rogozin.data,csuros.scenarios}. 
The ecdysozoan ancestor has an even higher estimated intron density 
(80\% of human density) than 
the otherwise quite generous estimates (about 70\%) of \shortciteN{RoyGilbert.earlygenes}, 
which is mostly due to the inclusion of the relatively intron-rich 
honeybee genes \shortcite{honeybee}.
Intron density in the bilaterian ancestor is 
estimated to be almost as high as in humans (94\%), 
agreeing with estimates of \shortciteN{RoyGilbert.earlygenes}. 
Sequences of a handful of intron-rich genes from 
the marine annelid {\em Platynereis dumerilii}
have already indicated 
that the bilaterian ancestor's genome was at least two-thirds 
as intron-rich as the vertebrate genomes
even by conservative estimates~\shortcite{introns.dumerilii,Roy.dumerilii}.

\begin{figure}
\centerline{\includegraphics[width=\columnwidth]{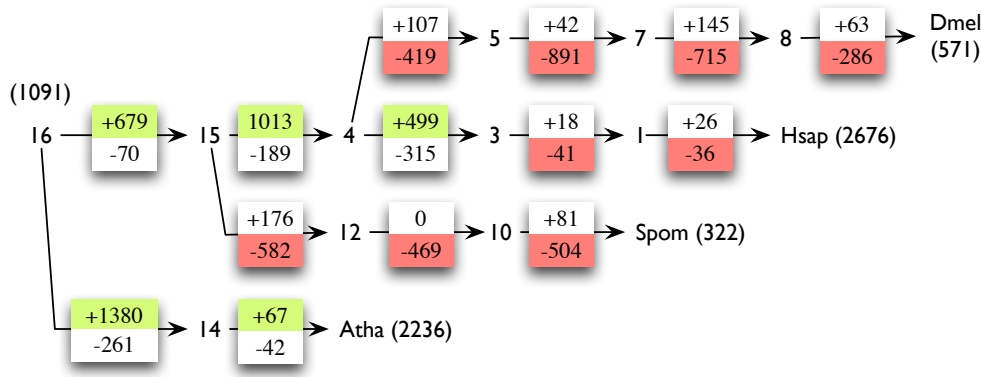}}
\caption{Intron gains and losses in a few evolutionary paths. 
	Estimated and actual intron counts are in parentheses.}\label{fig:changes}
\end{figure}

Introns are for the most part lost on the branches. Figure~\ref{fig:changes} 
shows the estimated changes in a few lineages.  
Intron evolution has been much slower in the vertebrate 
lineage than in most other lineages: in 
more than 380 million years since the divergence 
with fishes, only about 3\% of our introns got lost. 
Fungi, for example, massively
trimmed their introns in many lineages.
A notable exception is {\em C.~neoformans}, which seems to have gained 
introns, but that assessment may change if another
basidiomycete genome becomes available besides
the relatively intron-poor {\em U.~maydis}. 

\section{Conclusion}
We presented a novel alignment technique for 
establishing intron orthology, and a likelihood 
framework in which intron evolutionary events can be quantified. 
We described a compression method for the evaluation 
of the likelihood function, which has been extremely valuable 
in practice. We also showed that 
the compression leads to sublinear running times
for likelihood evaluation.

We illustrated our methods for analyzing intron evolution with a 
large and diverse set of eukaryotic organisms. 
The data set is more comprehensive 
than any used in other studies published to this day. 
The data indicate that ancestral eukaryotic genomes
were more intron-rich than previous studies suggested. 

Many circumstances influence intron loss \shortcite{Jeffares.gainloss,RoyGilbert.review},
and realistic likelihood models need to 
introduce rate variation \shortcite{Carmel.EM}. 
Usual rate variation models~\shortcite{Felsenstein} 
entail multiple evaluations of the likelihood function, 
and, thus, underline the importance of computational efficiency.
We believe that the proposed methods will help to produce 
and analyze large data sets even within complicated 
likelihood models.

\section*{Acknowledgment}
This research is supported by a grant from 
the Natural Sciences and Engineering Research Council of Canada.


\begin{thebibliography}{}

\bibitem[\protect\citeauthoryear{Adachi and Hasegawa}{Adachi and
  Hasegawa}{1995}]{MOLPHY}
Adachi, J. and M.~Hasegawa (1995).
\newblock {MOLPHY} version 2.3: programs for molecular phylogenetics based on
  maximum likelihood.
\newblock Volume~28 of {\em Computer Science Monographs}, pp.\  1--150. Tokyo,
  Japan: Institute of Statistical Mathematics.

\bibitem[\protect\citeauthoryear{Aldous}{Aldous}{2001}]{Aldous.beta}
Aldous, D. (2001).
\newblock Stochastic models and descriptive statistics for phylogenetic trees,
  from {Y}ule to today.
\newblock {\em Statistical Science\/}~{\em 16}, 23--34.

\bibitem[\protect\citeauthoryear{Altschul, Madden, Sch{\"a}ffer, Zhang, Zhang,
  Miller, and Lipman}{Altschul et~al.}{1997}]{BLAST}
Altschul, S.~F., T.~L. Madden, A.~A. Sch{\"a}ffer, J.~Zhang, Z.~Zhang,
  W.~Miller, and D.~J. Lipman (1997).
\newblock Gapped {BLAST} and {PSI-BLAST}: a new generation of protein database
  search programs.
\newblock {\em Nucleic Acids Research\/}~{\em 25\/}(17), 3389--3402.

\bibitem[\protect\citeauthoryear{Bieri et~al.}{Bieri
  et~al.}{2007}]{WormBase2007}
Bieri, T. et~al. (2007).
\newblock {WormBase}: new content and better access.
\newblock {\em Nucleic Acids Research\/}~{\em 35\/}(suppl\_1), D506--510.

\bibitem[\protect\citeauthoryear{Blum and Fran{\c c}ois}{Blum and Fran{\c
  c}ois}{2005}]{BlumFrancois.Sackin}
Blum, M. G.~B. and O.~Fran{\c c}ois (2005).
\newblock On statistical tests of phylogenetic tree imbalance: The {S}ackin and
  other indices revisited.
\newblock {\em Mathematical Biosciences\/}~{\em 195}, 141--153.

\bibitem[\protect\citeauthoryear{Carmel, Rogozin, Wolf, and Koonin}{Carmel
  et~al.}{2005}]{Carmel.EM}
Carmel, L., I.~B. Rogozin, Y.~I. Wolf, and E.~V. Koonin (2005).
\newblock An expectation-maximization algorithm for analysis of evolution of
  exon-intron structure of eukaryotic genes.
\newblock See \citeN{RCG2005}, pp.\  35--46.

\bibitem[\protect\citeauthoryear{Collins and Penny}{Collins and
  Penny}{2005}]{CollinsPenny}
Collins, L. and D.~Penny (2005).
\newblock Complex spliceosomal organization ancestral to extant eukaryotes.
\newblock {\em Molecular Biology and Evolution\/}~{\em 22\/}(4), 1053--1066.

\bibitem[\protect\citeauthoryear{Coulombe-Huntington and
  Majewski}{Coulombe-Huntington and Majewski}{2007}]{Coulombe.mammals}
Coulombe-Huntington, J. and J.~Majewski (2007).
\newblock Characterization of intron loss events in mammals.
\newblock {\em Genome Research\/}~{\em 17\/}(1), 23--32.

\bibitem[\protect\citeauthoryear{Cs{\H u}r\"os}{Cs{\H
  u}r\"os}{2004}]{segment-sets}
Cs{\H u}r\"os, M. (2004).
\newblock Maximum-scoring segment sets.
\newblock {\em IEEE/ACM Transactions on Computational Biology and
  Bioinformatics\/}~{\em 1\/}(4), 139--150.

\bibitem[\protect\citeauthoryear{Cs{\H u}r\"os}{Cs{\H
  u}r\"os}{2005}]{csuros.scenarios}
Cs{\H u}r\"os, M. (2005).
\newblock Likely scenarios of intron evolution.
\newblock See \citeN{RCG2005}, pp.\  47--60.
\newblock DOI:10.1007/11554714\_5.

\bibitem[\protect\citeauthoryear{Devroye}{Devroye}{1991}]{Devroye.subtrees.nor%
mal}
Devroye, L. (1991).
\newblock Limit laws for local counters in random binary search trees.
\newblock {\em Random Structures and Algorithms\/}~{\em 2\/}(1), 303--315.

\bibitem[\protect\citeauthoryear{Durbin, Eddy, Krogh, and Mitchison}{Durbin
  et~al.}{1998}]{Durbin}
Durbin, R., S.~R. Eddy, A.~Krogh, and G.~Mitchison (1998).
\newblock {\em Biological Sequence Analysis: Probabilistic Models of Proteins
  and Nucleic Acids}.
\newblock UK: Cambridge University Press.

\bibitem[\protect\citeauthoryear{Edgar}{Edgar}{2004}]{MUSCLE}
Edgar, R.~C. (2004).
\newblock {MUSCLE}: multiple sequence alignment with high accuracy and high
  throughput.
\newblock {\em Nucleic Acids Research\/}~{\em 32\/}(5), 1792--1797.

\bibitem[\protect\citeauthoryear{Felsenstein}{Felsenstein}{1981}]{Felsenstein.%
ML}
Felsenstein, J. (1981).
\newblock Evolutionary trees from {DNA} sequences: A maximum likelihood
  approach.
\newblock {\em Journal of Molecular Evolution\/}~{\em 17}, 368--376.

\bibitem[\protect\citeauthoryear{Felsenstein}{Felsenstein}{1992}]{Felsenstein.%
restml}
Felsenstein, J. (1992).
\newblock Phylogenies from restriction sites, a maximum likelihood approach.
\newblock {\em Evolution\/}~{\em 46}, 159--173.

\bibitem[\protect\citeauthoryear{Felsenstein}{Felsenstein}{2004}]{Felsenstein}
Felsenstein, J. (2004).
\newblock {\em Inferring Pylogenies}.
\newblock Sunderland, Mass.: Sinauer Associates.

\bibitem[\protect\citeauthoryear{Friedman, Ninio, Pupko, Pe'er, and
  Pupko}{Friedman et~al.}{2002}]{SEMPHY}
Friedman, N., M.~Ninio, T.~Pupko, I.~Pe'er, and T.~Pupko (2002).
\newblock A structural {EM} algorithm for phylogenetic inference.
\newblock {\em Journal of Computational Biology\/}~{\em 9\/}(2), 331--353.

\bibitem[\protect\citeauthoryear{Guindon and Gascuel}{Guindon and
  Gascuel}{2003}]{PhyML}
Guindon, S. and O.~Gascuel (2003).
\newblock A simple, fast, and aaccurate accurate algorithm to estimate large
  phylogenies by maximum likelihood.
\newblock {\em Systematic Biology\/}~{\em 52\/}(5), 696--704.

\bibitem[\protect\citeauthoryear{Harding}{Harding}{1971}]{Harding}
Harding, E. (1971).
\newblock The probabilities of rooted tree-shapes generated by random
  bifurcation.
\newblock {\em Advances in Applied Probability\/}~{\em 3}, 44--77.

\bibitem[\protect\citeauthoryear{Heard}{Heard}{1992}]{Heard.subtrees}
Heard, S.~B. (1992).
\newblock Patterns in tree balance among cladistic, phenetic, and randomly
  generated phylogenetic trees.
\newblock {\em Evolution\/}~{\em 46\/}(6), 1818--1826.

\bibitem[\protect\citeauthoryear{Hubbard et~al.}{Hubbard
  et~al.}{2007}]{Ensembl2007}
Hubbard, T. J.~P. et~al. (2007).
\newblock Ensembl 2007.
\newblock {\em Nucleic Acids Research\/}~{\em 35\/}(suppl\_1), D610--617.

\bibitem[\protect\citeauthoryear{IHBSC}{IHBSC}{2006}]{honeybee}
IHBSC (2006).
\newblock Insights into social insects from the genome of the honey bee {A}pis
  mellifera.
\newblock {\em Nature\/}~{\em 443}, 931--949.

\bibitem[\protect\citeauthoryear{Jeffares, Mourier, and Penny}{Jeffares
  et~al.}{2006}]{Jeffares.gainloss}
Jeffares, D.~C., T.~Mourier, and D.~Penny (2006).
\newblock The biology of intron gain and loss.
\newblock {\em Trends in Genetics\/}~{\em 22\/}(1), 16--22.

\bibitem[\protect\citeauthoryear{Kececioglu and Zhang}{Kececioglu and
  Zhang}{1998}]{KececiogluZhang}
Kececioglu, J. and W.~Zhang (1998).
\newblock Aligning alignments.
\newblock In M.~Farach-Colton (Ed.), {\em Proc.\ CPM}, Volume 1448 of {\em
  LNCS}, pp.\  189--208. Springer.

\bibitem[\protect\citeauthoryear{Ma, Wang, and Zhang}{Ma
  et~al.}{2003}]{aliali.NP}
Ma, B., Z.~Wang, and K.~Zhang (2003).
\newblock Alignment between two multiple alignments.
\newblock In R.~A. Baeza-Yates, E.~Ch{\'a}vez, and M.~Crochemore (Eds.), {\em
  Proc.\ Combinatorial Pattern Matching (CPM)}, Volume 2676 of {\em LNCS}, pp.\
   254--265. Springer.

\bibitem[\protect\citeauthoryear{Marchler-Bauer et~al.}{Marchler-Bauer
  et~al.}{2007}]{CDD2007}
Marchler-Bauer, A. et~al. (2007).
\newblock {CDD}: a conserved domain database for interactive domain family
  analysis.
\newblock {\em Nucleic Acids Research\/}~{\em 35\/}(suppl\_1), D237--240.

\bibitem[\protect\citeauthoryear{McDiarmid}{McDiarmid}{1989}]{McDiarmid}
McDiarmid, C. (1989).
\newblock On the method of bounded differences.
\newblock In J.~Siemons (Ed.), {\em Surveys in Combinatorics}, pp.\  148--184.
  Cambridge University Press.

\bibitem[\protect\citeauthoryear{McKenzie and Steel}{McKenzie and
  Steel}{2000}]{McKenzieSteel.cherries}
McKenzie, A. and M.~Steel (2000).
\newblock Distributions of cherries for two models of trees.
\newblock {\em Mathematical Biosciences\/}~{\em 164}, 81--92.

\bibitem[\protect\citeauthoryear{McLysaght and Huson}{McLysaght and
  Huson}{2005}]{RCG2005}
McLysaght, A. and D.~Huson (Eds.) (2005).
\newblock {\em Proc.\ RECOMB Satellite Workshop on Comparative Genomics},
  Volume 3678 of {\em LNCS}.
\newblock Springer-Verlag.

\bibitem[\protect\citeauthoryear{M{\"u}ller, Spang, and Vingron}{M{\"u}ller
  et~al.}{2002}]{VTML}
M{\"u}ller, T., R.~Spang, and M.~Vingron (2002).
\newblock Estimating amino acid substitution models: a comparison of
  {D}ayhoff's estimator, the resolvent approach and a maximum likelihood
  method.
\newblock {\em Molecular Biology and Evolution\/}~{\em 19\/}(1), 8--13.

\bibitem[\protect\citeauthoryear{Nguyen, Yoshihama, and Kenmochi}{Nguyen
  et~al.}{2005}]{Nguyen.introns}
Nguyen, H.~D., M.~Yoshihama, and N.~Kenmochi (2005).
\newblock New maximum likelihood estimators for eukaryotic intron evolution.
\newblock {\em PLoS Computational Biology\/}~{\em 1\/}(7), e79.

\bibitem[\protect\citeauthoryear{Nielsen, Friendman, Birren, Burge, and
  Galagan}{Nielsen et~al.}{2004}]{introns.fungi}
Nielsen, C.~B., B.~Friendman, B.~Birren, C.~B. Burge, and J.~E. Galagan (2004).
\newblock Patterns of intron gain and loss in fungi.
\newblock {\em PLoS Biology\/}~{\em 2\/}(12), e422.

\bibitem[\protect\citeauthoryear{Nixon, Wang, Morrison, McArthur, Sogin,
  Loftus, and Samuelson}{Nixon et~al.}{2002}]{introns.giardia}
Nixon, J. E.~J., A.~Wang, H.~G. Morrison, A.~G. McArthur, M.~L. Sogin, B.~J.
  Loftus, and J.~Samuelson (2002).
\newblock A spliceosomal intron in {G}iardia lamblia.
\newblock {\em Proceedings of the National Academy of Sciences of the
  USA\/}~{\em 99\/}(6), 3701--3705.

\bibitem[\protect\citeauthoryear{Press, Teukolsky, Vetterling, and
  Flannery}{Press et~al.}{1997}]{NR}
Press, W.~H., S.~A. Teukolsky, W.~V. Vetterling, and B.~P. Flannery (1997).
\newblock {\em Numerical Recipes in C: The Art of Scientific Computing\/}
  (Second ed.).
\newblock Cambridge UniversIty Press.

\bibitem[\protect\citeauthoryear{Pruitt, Tatusova, and Maglott}{Pruitt
  et~al.}{2007}]{RefSeq2007}
Pruitt, K.~D., T.~Tatusova, and D.~R. Maglott (2007).
\newblock {NCBI} reference sequences ({RefSeq}): a curated non-redundant
  sequence database of genomes, transcripts and proteins.
\newblock {\em Nucleic Acids Research\/}~{\em 35\/}(suppl\_1), D61--65.

\bibitem[\protect\citeauthoryear{Raible, Tessmar-Raible, Osoegawa, Wincker,
  Jubin, Balavoine, Ferrier, Benes, de~Jong, Weissenbach, Bork, and
  Arendt}{Raible et~al.}{2005}]{introns.dumerilii}
Raible, F., K.~Tessmar-Raible, K.~Osoegawa, P.~Wincker, C.~Jubin, G.~Balavoine,
  D.~Ferrier, V.~Benes, P.~de~Jong, J.~Weissenbach, P.~Bork, and D.~Arendt
  (2005).
\newblock Vertebrate-type intron-rich genes in the marine annelid {P}latynereis
  dumerilii.
\newblock {\em Science\/}~{\em 310}, 1325--1326.

\bibitem[\protect\citeauthoryear{Rogozin, Sverdlov, Babenko, and
  Koonin}{Rogozin et~al.}{2005}]{Rogozin.methods}
Rogozin, I.~B., A.~V. Sverdlov, V.~N. Babenko, and E.~V. Koonin (2005).
\newblock Analysis of evolution of exon-intron structure of eukaryotic genes.
\newblock {\em Briefings in Bioinformatics\/}~{\em 6\/}(2), 118--134.

\bibitem[\protect\citeauthoryear{Rogozin, Wolf, Sorokin, Mirkin, and
  Koonin}{Rogozin et~al.}{2003}]{Rogozin.data}
Rogozin, I.~B., Y.~I. Wolf, A.~V. Sorokin, B.~G. Mirkin, and E.~V. Koonin
  (2003).
\newblock Remarkable interkingdom conservation of intron positions and massive,
  lineage-specific intron loss and gain in eukaryotic evolution.
\newblock {\em Current Biology\/}~{\em 13}, 1512--1517.

\bibitem[\protect\citeauthoryear{Rosenberg}{Rosenberg}{2006}]{Rosenberg.subtre%
es}
Rosenberg, N.~A. (2006).
\newblock The mean and variance of $r$-pronged nodes and $r$-caterpillars in
  {Y}ule-generated genealogies.
\newblock {\em Annals of Combinatorics\/}~{\em 10}, 129--146.

\bibitem[\protect\citeauthoryear{Ross}{Ross}{1996}]{Ross}
Ross, S.~M. (1996).
\newblock {\em Stochastic Processes\/} (Second ed.).
\newblock Wiley \&\ Sons.

\bibitem[\protect\citeauthoryear{Roy}{Roy}{2006}]{Roy.dumerilii}
Roy, S.~W. (2006).
\newblock Intron-rich ancestors.
\newblock {\em Trends in Genetics\/}~{\em 22\/}(9), 468--471.

\bibitem[\protect\citeauthoryear{Roy and Gilbert}{Roy and
  Gilbert}{2005}]{RoyGilbert.earlygenes}
Roy, S.~W. and W.~Gilbert (2005).
\newblock Complex early genes.
\newblock {\em Proceedings of the National Academy of Sciences of the
  USA\/}~{\em 102\/}(6), 1986--1991.

\bibitem[\protect\citeauthoryear{Roy and Gilbert}{Roy and
  Gilbert}{2006}]{RoyGilbert.review}
Roy, S.~W. and W.~Gilbert (2006).
\newblock The evolution of spliceosomal introns: patterns, puzzles and
  progress.
\newblock {\em Nature Reviews Genetics\/}~{\em 7}, 211--221.

\bibitem[\protect\citeauthoryear{Roy and Penny}{Roy and
  Penny}{2006}]{Roy.apicomplexa}
Roy, S.~W. and D.~Penny (2006).
\newblock Large-scale intron conservation and order-of-magnitude variation in
  intron loss/gain rates in apicomplexan evolution.
\newblock {\em Genome Research\/}~{\em 16\/}(10), 1270--1275.

\bibitem[\protect\citeauthoryear{Roy and Penny}{Roy and
  Penny}{2007}]{Roy.plants}
Roy, S.~W. and D.~Penny (2007).
\newblock Patterns of intron loss and gain in plants: Intron loss-dominated
  evolution and genome-wide comparison of {O}.~sativa and {A}.~thaliana.
\newblock {\em Molecular Biology and Evolution\/}~{\em 24\/}(1), 171--181.

\bibitem[\protect\citeauthoryear{Stamatakis, Ludwig, Meier, and
  Wolf}{Stamatakis et~al.}{2002}]{AxML}
Stamatakis, A.~P., T.~Ludwig, H.~Meier, and M.~J. Wolf (2002).
\newblock {AxML}: A fast program for sequential and parallel phylogenetic tree
  calculations based on the maximum likelihood method.
\newblock In {\em Proc.\ IEEE Computer Society Bioinformatics Conference
  (CSB)}, pp.\  21--28.

\bibitem[\protect\citeauthoryear{Steel}{Steel}{1994}]{Steel.markov}
Steel, M.~A. (1994).
\newblock Recovering a tree from the leaf colourations it generates under a
  {Markov} model.
\newblock {\em Applied Mathematics Letters\/}~{\em 7}, 19--24.

\bibitem[\protect\citeauthoryear{Sverdlov, Cs\H{u}r\"os, Rogozin, and
  Koonin}{Sverdlov et~al.}{2007}]{Sverdlov.ancientdup}
Sverdlov, A.~V., M.~Cs\H{u}r\"os, I.~B. Rogozin, and E.~V. Koonin (2007).
\newblock A glimpse of a putative pre-intron phase of eukaryotic evolution.
\newblock {\em Trends in Genetics\/}.
\newblock In press. DOI:~/10.1016/j.tig.2007.01.001.

\bibitem[\protect\citeauthoryear{Sverdlov, Rogozin, Babenko, and
  Koonin}{Sverdlov et~al.}{2005}]{Sverdlov.parallelgain}
Sverdlov, A.~V., I.~B. Rogozin, V.~N. Babenko, and E.~V. Koonin (2005).
\newblock Conservation versus parallel gains in intron evolution.
\newblock {\em Nucleic Acids Research\/}~{\em 33\/}(6), 1741--1748.

\bibitem[\protect\citeauthoryear{Tatusov et~al.}{Tatusov
  et~al.}{2003}]{COG.new}
Tatusov, R.~L. et~al. (2003).
\newblock The {COG} database: an updated version includes eukaryotes.
\newblock {\em BMC Bioinformatics\/}~{\em 4}, 441.

\bibitem[\protect\citeauthoryear{Va{\v n}\'a{\v c}ov\'a, Yan, Carlton, and
  Johnson}{Va{\v n}\'a{\v c}ov\'a et~al.}{2005}]{introns.trichomonas}
Va{\v n}\'a{\v c}ov\'a, {\v S}., W.~Yan, J.~M. Carlton, and P.~J. Johnson
  (2005).
\newblock Spliceosomal introns in the deep-branching eukaryote {T}richomonas
  vaginalis.
\newblock {\em Proceedings of the National Academy of Sciences of the
  USA\/}~{\em 102\/}(12), 4430--4435.

\bibitem[\protect\citeauthoryear{Zhang, Berman, Wiehe, and Miller}{Zhang
  et~al.}{1999}]{postprocessing}
Zhang, Z., P.~Berman, T.~Wiehe, and W.~Miller (1999).
\newblock Post-processing long pairwise alignments.
\newblock {\em Bioinformatics\/}~{\em 15\/}(12), 1012--1019.

\end{thebibliography}

\end{document}